\documentclass[twocolumn]{aastex61}

\received{August 4, 2017}
\revised{October 20, 2017}
\accepted{\today}
\submitjournal{AJ}

\shorttitle{RR Lyrae stars in Fornax}
\shortauthors{Karczmarek et al.}

\begin{document}

\title{THE ARAUCARIA PROJECT: THE DISTANCE TO THE FORNAX DWARF GALAXY FROM NEAR-INFRARED PHOTOMETRY OF RR~LYRAE STARS
\footnote{Based on data collected with the VLT/HAWK-I instrument at ESO Paranal Observatory, Chile, as a part of a programme 082.D-0123(B).}}

\correspondingauthor{Paulina Karczmarek}
\email{pkarczmarek@astrouw.edu.pl}

\author[0000-0002-0136-0046]{Paulina Karczmarek}
\affiliation{Warsaw University Observatory, Al. Ujazdowskie 4, 00-478, Warsaw, Poland}

\author{Grzegorz Pietrzy{\'n}ski}
\affiliation{Nicolaus Copernicus Astronomical Center, Bartycka 18, 00-716, Warsaw, Poland}
\affiliation{Departamento de Astronomia, Universidad de Concepci{\'o}n, Casilla 160-C, Concepci{\'o}n, Chile}

\author{Marek G{\'o}rski}
\affiliation{Departamento de Astronomia, Universidad de Concepci{\'o}n, Casilla 160-C, Concepci{\'o}n, Chile}
\affiliation{Millenium Institute of Astrophysics, Av. Vicu{\~n}a Mackenna, 4860 Santiago, Chile}

\author{Wolfgang Gieren}
\affiliation{Departamento de Astronomia, Universidad de Concepci{\'o}n, Casilla 160-C, Concepci{\'o}n, Chile}
\affiliation{Millenium Institute of Astrophysics, Av. Vicu{\~n}a Mackenna, 4860 Santiago, Chile}

\author{David Bersier}
\affiliation{Astrophysics Research Institute, Liverpool Science Park, 146 Brownlow Hill, Liverpool L3 5RF, United Kingdom}

\begin{abstract}
We have obtained single-phase near-infrared (NIR) magnitudes in the $J$- and $K$-bands for 77 RR~Lyrae (RRL) stars in the Fornax Dwarf Spheroidal Galaxy. We have used different theoretical and empirical NIR period-luminosity-metallicity calibrations for RRL stars to derive their absolute magnitudes, and found a true, reddening-corrected distance modulus of $20.818\pm 0.015 \mbox{ (statistical)} \pm 0.116 \mbox{ (systematic)}$ mag. This value is in excellent agreement with the results obtained within the Araucaria Project from the NIR photometry of red clump stars ($20.858 \pm 0.013$ mag), the tip of the red giant branch ($20.84 \pm 0.04 \pm 0.14$ mag), as well as with other independent distance determinations to this galaxy. The effect of metallicity and reddening is substantially reduced in the NIR domain, making this method a robust tool for accurate distance determination at the 5 percent level. This precision is expected to reach the level of 3 percent once the zero points of distance calibrations are refined thanks to the Gaia mission. NIR period-luminosity-metallicity relations of RRL stars are particularly useful for distance determinations to galaxies and globular clusters up to 300 kpc, that lack young standard candles, like Cepheids.
\end{abstract}

\keywords{galaxies: distances and redshift --- galaxies: individual (Fornax) --- infrared: stars --- stars: variables: RR Lyrae}

\section{Introduction}

The main focus of the Araucaria Project is to improve the calibrations of the distance scale, with the use of major distance indicators in several nearby galaxies \citep[e.g.][]{gieren05, pietrzynski13eas}. Such a study is capable of revealing the impact of metallicity or/and age on various standard candles. Understanding and minimizing the uncertainties of distance determinations in the local scale improves the precision and accuracy of secondary distance indicators, ultimately leading to better determination of the Hubble parameter. Over the past 15 years of the Araucaria Project, distances to nearby galaxies were determined with the use of the red clump stars \citep{pietrzynski02}, the tip of the red giant branch \citep{pietrzynski09,gorski11}, late-type eclipsing binaries \citep{pietrzynski13nat,graczyk14}, blue supergiants \citep{urbaneja08}, classical Cepheids \citep[e.g.][]{gieren13,nardetto14}, type II Cepheids \citep{ciechanowska10}, and RR Lyrae stars \citep{pietrzynski08,szewczyk08,szewczyk09,karczmarek15}.

Distance determinations can be significantly improved in the near-infrared (NIR) spectral range, which has been long known to be markedly less affected by interstellar extinction and metallicity \citep{mcgonegal83}. This is especially useful in the case of RR~Lyrae (RRL) stars, which throughout the years of theoretical and empirical studies have proven to be excellent standard candles in the NIR domain, providing distance results which are superior to optical studies \citep[e.g.][]{longmore86,bono03}. The NIR amplitudes of RRLs are a factor of 2--3 smaller than in the optical bands, giving the NIR full-amplitude of only 0.4~mag \citep[e.g.][]{storm92,marconi03}. Therefore, mean NIR magnitudes can be approximated by random-phase magnitudes obtained in a modest number of observations, as long as a significant RRL sample is concerned. Yet the biggest advantage of the NIR observations of RRL stars is that they follow the period-luminosity relation \citep[PL,][]{longmore86,fernley87}. This feature was studied extensively by \citet{bono01}, who put the first theoretical constrains on $K$-band PL relation of RRLs based on nonlinear convective pulsation models. The continuation of theoretical studies of RRL stars in NIR wavebands, carried out by \citet{bono03} and \citet{catelan04}, yielded the period-luminosity-metallicity (PLZ) relation in the NIR domain, showing that the effect of metallicity on the luminosity of RRLs in the NIR domain is noticeably smaller as compared to the visual one. Indeed, the metallicity term enters the PLZ relation with the coefficient of at most 0.23~mag\,dex$^{-1}$ in $K$-band \citep{bono03}, while in the $V$-band the metallicity coefficient reaches 0.3~mag\,dex$^{-1}$ \citep{dicriscienzo04}. Empirical PLZ relations followed the theoretical studies and showed the metallicity coefficient to be even smaller \citep[0.08~mag\,dex$^{-1}$,][]{sollima08}, but no general consensus has been reached yet on the value of the coefficient of the metallicity term.

Within the Araucaria Project, we repeatedly showed that the NIR PLZ relation of RRL stars is a superior and reliable tool, by determining distance moduli to a number of galaxies in the Local Group: the Sculptor dwarf galaxy \citep{pietrzynski08}, the Magellanic Clouds \citep{szewczyk08,szewczyk09}, and the Carina dwarf galaxy \citep{karczmarek15}, with precision at the 5\% level or better. In this paper, we use $J$- and $K$-band PLZ relations for RRL stars to deliver the distance modulus to the Fornax galaxy.

The Fornax dSph Galaxy is an extraordinary stellar system, not only because it is one of the most massive satellites of the Milky Way, but also due to five distinctive globular clusters that reside in its body. Numerous studies of Fornax continuously add new findings to its complex picture: 
large internal metallicity spread $-2.5 \leq \mbox{[Fe/H]} \leq 0.0$ \citep{tolstoy01,pont04,battaglia06}, the presence of three stellar populations \citep{battaglia06}, the off-center overdense structure of debatable origin \citep{coleman04,deboer13}, the mean period of sub-type RRab stars falling in between average periods for Oosterhoff classes I and II \citep{bersier02}.
These diverse and inconclusive results amplify further interest in the Fornax galaxy. Precise and accurate determination of Fornax distance is therefore crucial in numerous distance-based studies, like the analysis of the color-magnitude diagram (CMD), as it reconciles the model with the observed CMD, and therefore allows for bias-free determination of the metallicity and age of stellar populations \citep{deboer16}.

The Fornax galaxy has been a target for many distance determination techniques in the optical and infrared domains. Distances were calculated based on the field stars as well as on the stars in each of its five globular clusters. Studies that involved field stars of the horizontal branch \citep{irwin95,saviane00,rizzi07}, the tip of the red giant branch \citep{saviane00,bersier00,gullieuszik07,pietrzynski09}, red clump stars \citep{bersier00,pietrzynski03,rizzi07}, the CMD fitting \citep{weisz14}, and RRL stars \citep{greco05astroph,mcnamara11} in the visual and infrared domains are summarized in Table \ref{tab:fordist}.
In this paper we complement existing distance determinations to the Fornax galaxy with a distance modulus derived from PLZ relations of field RRL stars in the $J$- and $K$-band. Taking advantage of the NIR magnitudes, which are only slightly affected by the metallicity and reddening, we achieve a precision at the 5\% level.

\begin{deluxetable*}{LcCc}
\tablewidth{0pc}
\tablecaption{
\label{tab:fordist}
Selected determinations of true distance moduli to the Fornax dSph Galaxy, obtained with different stellar indicators in the field of the galaxy, in the optical and near-infrared domains. If available, statistical and systematic errors were included.}
\tablehead{
\colhead{Distance modulus} &
\colhead{Method\tablenotemark{a}} &
\colhead{Filter} &
\colhead{Ref.} \\
\colhead{(mag)} & \colhead{} & \colhead{} & \colhead{}
}
\startdata
20.76 \pm 0.10	 		& RGB 	& V	& \citet{buonanno99} \\
20.79					& CMD	& F814W, F555W & \citet{weisz14} \\
20.89 \pm 0.18\tablenotemark{b}	& CS & W_{JK} & \citet{huxor15} \\
20.70 \pm 0.12			& TRGB 	& V, I & \citet{saviane00} \\
20.65 \pm 0.11			& TRGB	& I	& \citet{bersier00} \\
20.75 \pm 0.19			& TRGB	& K	& \citet{gullieuszik07} \\
20.71 \pm 0.07 			& TRGB	& I & \citet{rizzi07} \\
20.84 \pm 0.03 \pm 0.12	& TRGB	& J	& \citet{pietrzynski09} \\
20.84 \pm 0.04 \pm 0.14	& TRGB	& K	& \citet{pietrzynski09} \\
20.66					& RC	& I	& \citet{bersier00} \\
20.858 \pm 0.013		& RC	& K	& \citet{pietrzynski03} \\
20.74 \pm 0.11			& RC 	& K	& \citet{gullieuszik07} \\
20.72 \pm 0.04 			& RC 	& I & \citet{rizzi07} \\
20.40 \pm 0.14 			& HB 	& V & \citet{irwin95}\\
20.76 \pm 0.04			& HB	& V	&  \citet{saviane00} \\ 
20.72 \pm 0.06 			& HB 	& I & \citet{rizzi07} \\
20.70 \pm 0.02			& SXP	& V	& \citet{poretti08} \\
20.93 \pm 0.07			& SXP	& V	& \citet{mcnamara11} \\
20.72 \pm 0.10			& RRL	& V & \citet{greco05astroph} \\
20.90 \pm 0.05			& RRL	& V	& \citet{mcnamara11} \\
20.818\pm 0.015 \pm 0.116	& RRL	& J, K	& This paper \\
\enddata
\tablenotetext{a}{Distance determined from: comparison of Red Giant Branch (RGB) stars from Fornax and M5, color-magnitude diagram (CMD), Carbon Stars (CS), the Tip of Red Giant Branch (TRGB), Red Clump stars (RC), Horizontal Branch stars (HB), SX Phoenicis variables (SXP),  RR~Lyrae variables (RRL).}
\tablenotetext{b}{Value derived for the purpose of this paper by averaging over distance moduli of 18 carbon stars.}
\end{deluxetable*}

\section{Observations, data calibration \\and reduction}

\begin{figure}
\includegraphics[width=\columnwidth]{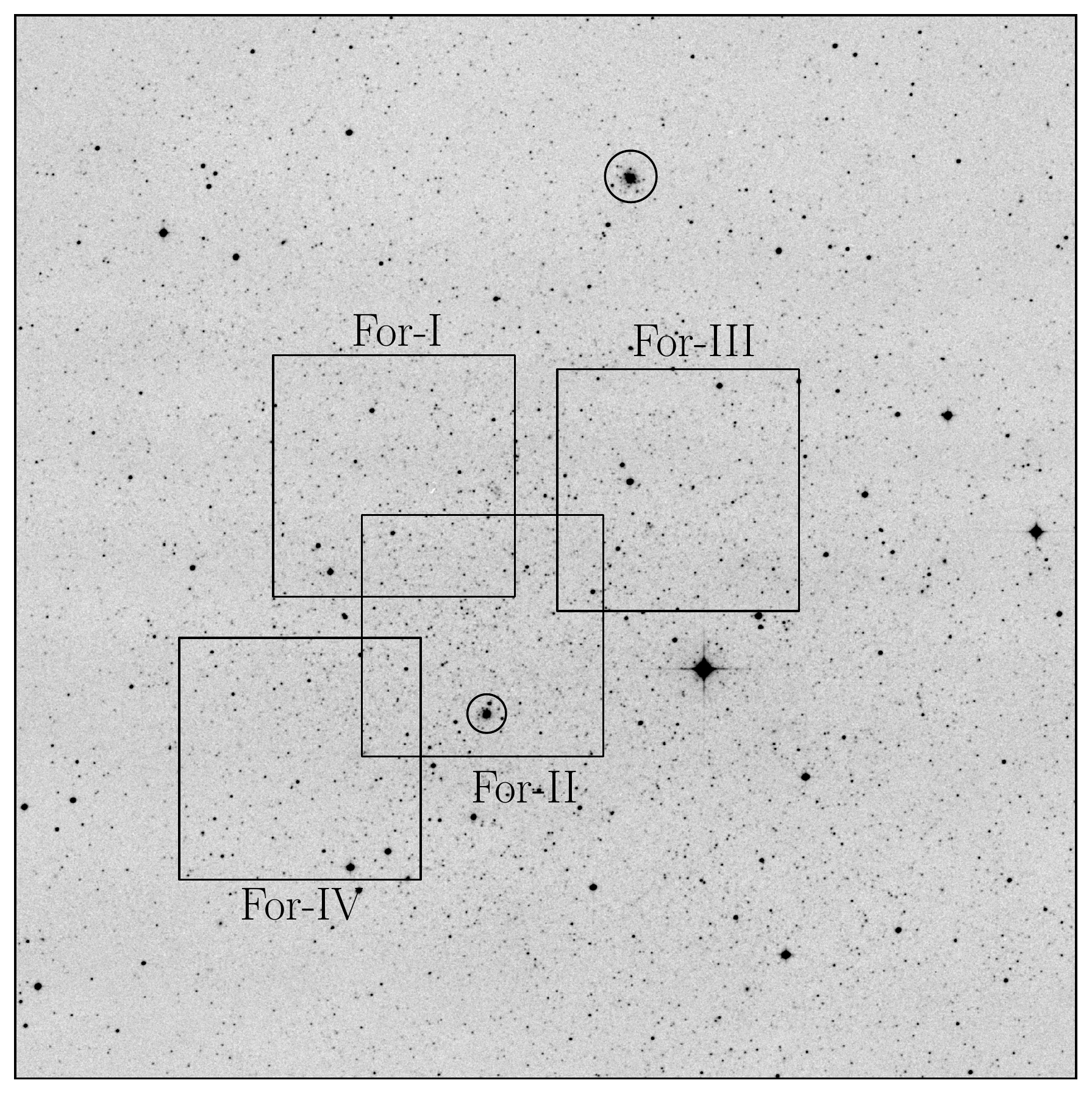}
\caption{
Four $7.5' \times 7.5'$ VLT/HAWK-I fields cover the north-east part of the Fornax galaxy, as marked on this $30' \times 30'$ DSS-2 infrared plate. North is up and east is to the left. Globular clusters are marked with circles: Fornax 4 lies inside the For-II field and near the center of Fornax galaxy, Fornax 3 (NGC 1049) lies above HAWK-I fields, on the top of the figure. Radii of circles are exaggerated and do not correspond to clusters' tidal radii.}
\label{fig:field}
\end{figure}

NIR observations were conducted between 2008 October 7 and 2008 October 23 with the High Acuity Wide-field K-band Imager (HAWK-I) mounted in the Nasmyth focus of UT4/VLT ESO 8\,m telescope at Paranal, Chile. The dates and coordinates of target fields are given in Table \ref{table:obs}. The locations of four target fields in the Fornax galaxy (Fig. \ref{fig:field}) were purposely chosen to overlap fields from the previous visual observations of \citet{bersier02}. The HAWK-I field of view of four $2048 \times 2048$ pixels detectors was about $7.5' \times 7.5'$ with a scale of $0.106'' \, \mathrm{pixel}^{-1}$. Detailed description of HAWK-I instrument can be found in \citet{kissler08}, and on the ESO website\footnote{http://www.eso.org/sci/facilities/paranal/instruments/hawki.html}. The fields were observed in $J$ and $K_{\rm s}$ (further denoted as $K$) wavebands, under photometric conditions (seeing range 0.4--0.7$''$). 
Our objects, having minimal brightness higher than 20.8 mag ($J$) and 20.6 mag ($K$), were well above the limiting HAWK-I magnitudes, i.e. 23.9 mag and 22.3 mag in $J$- and $K$-band, respectively \citep{kissler08}.
In order to account for frequent sky-level variations, which are especially strong in the NIR domain, the observations were carried out in the jitter mode, meaning that the consecutive exposures of a given field were randomly shifted within the radius of $20''$ with respect to the initial position. Exposure times of 60 seconds per jitter were fixed for the $J$- and $K$-band, resulting in total integration times of 18 and 26 minutes per field, respectively.

\begin{deluxetable}{cccc}
\tablewidth{0pc}
\tablecaption{
\label{table:obs}
The journal of observations of the target fields.}
\tablehead{
\colhead{Field} &
\colhead{R.A.}  &
\colhead{Dec.} &
\colhead{Date of} \\
\colhead{name}  &
\colhead{(J2000)} &
\colhead{(J2000)} &
\colhead{observation}
}
\startdata
For-I   & 02:40:22.4 & -34:24:52.5 & 2008 Oct 07, 09 \\
For-II  & 02:40:08.5 & -34:29:45.8 & 2008 Oct 23 \\
For-III & 02:39:39.6 & -34:25:06.3 & 2008 Oct 08 \\
For-IV  & 02:40:35.7 & -34:33:42.3 & 2008 Oct 09 \\
\enddata
\end{deluxetable}

\subsection{HAWK-I calibration pipeline}

The HAWK-I pipeline, a module of ESO Recipe Execution Tool (EsoRex), was utilized to execute complete reduction of our data. After basic calibration routines (dark correction, flat fielding and bad pixel correction), the subtraction of sky level was applied in a two-step process which included the masking of stars. Next, a refinement procedure minimized the effect of any distortions caused by atmospheric refraction or non-planar surface of detectors, and abolished the shifts in the images caused by the jittering. Lastly, frames were stacked into the final image. The procedure for standard stars was analogous, only consisted of fewer steps: basic calibration (dark correction, flat fielding), one-step sky subtraction, and distortion correction.

\subsection{Photometry}

Instrumental magnitudes were calculated with the pipeline developed in the course of the Araucaria Project. The point-spread function (PSF) photometry and aperture corrections were applied using DAOPHOT and ALLSTAR programs \citep{stetson87} in a way described by \citet{pietrzynski02}. In order to calibrate our photometric data to the standard photometric system, standard stars from the United Kingdom Infrared Telescope list \citep[UKIRT,][]{hawarden01} were observed at different air masses and spanning a broad range in color, bracketing the colors of RRL stars in the Fornax galaxy. The accuracy of the zero point of our photometry is about 0.02 mag. 

Due to insufficient number of standard stars observed 23 Oct 2008, an alternative approach was used in order to calibrate the magnitudes of stars in one field, For-II. We carried out follow-up observations of the same four HAWK-I fields with SOFI/NTT at ESO La Silla, Chile, executed reduction routine analogous to the routine described above for HAWK-I data, and performed the PSF photometry with aperture corrections in the exact same way as for HAWK-I data. The calibration of magnitudes of SOFI stars onto UKIRT photometric system was based on 10 standard stars, observed together with the target fields under photometric conditions and at different air masses. The matching algorithm paired stars from SOFI fields with remaining three HAWK-I fields: For-I, For-III, and For-IV, and found 184 common stars, which had already calibrated magnitudes from both instruments. Comparison of these magnitudes showed the consistency of the zero points at the level of 0.02~mag, and justified using SOFI calibrated magnitudes as a point of reference in order to calculate the systematical shift between the two data sets, and calibrate magnitudes of one remaining field For-II from HAWK-I data. 

\begin{deluxetable}{CCC}
\tablewidth{0pc}
\tablecaption{
\label{table:zp}
Difference of zero point calibration between 2MASS and SOFI, and our data obtained with HAWK-I.}
\tablehead{
\colhead{Filter} &
\colhead{2MASS -- HAWK-I}  &
\colhead{SOFI -- HAWK-I}\\
\colhead{} & \colhead{(mag)}  & \colhead{(mag)}}
\startdata
J  & -0.007 \pm 0.113	& 0.019 \pm 0.072 \\
K  & \phm{-}0.013 \pm 0.111	& 0.009 \pm 0.078 \\
\enddata
\end{deluxetable}

Additionally, we performed an independent brightness comparisons using data from Two Micron All Sky Survey (2MASS), in order to make a double check of our photometric zero point (Table \ref{table:zp}). We paired 99 brightest stars from HAWK-I dataset and 2MASS database, transformed the photometric system to 2MASS, and compared the magnitudes. The averaged zero-point differences from 2MASS and SOFI comparisons are also presented in Table \ref{table:zp}. These results firmly support the accuracy of the absolute calibration of HAWK-I data.

\subsection{Identification of variables}

\begin{figure}
\includegraphics[width=\columnwidth]{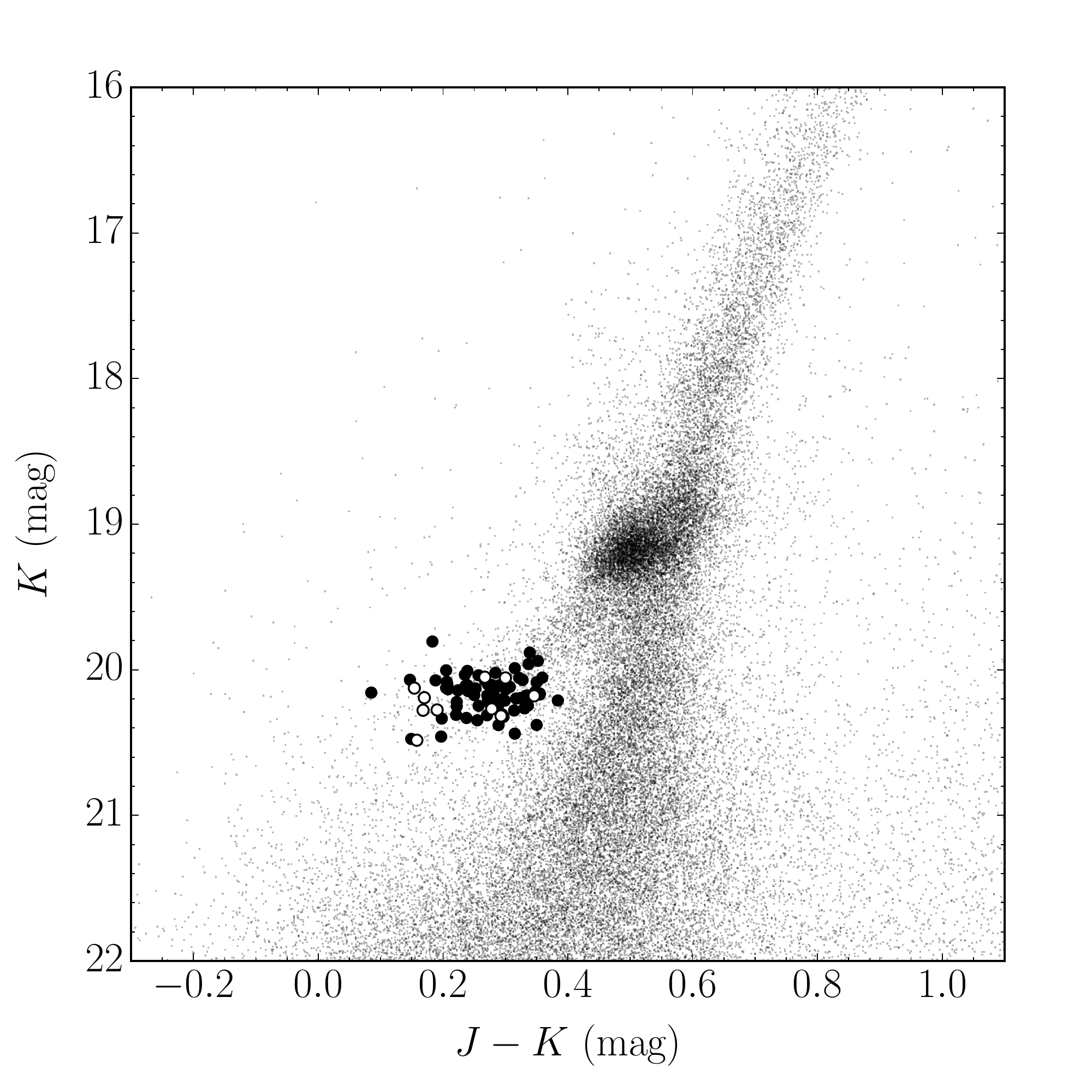}
\caption{
Infrared CMD showing identified RRab (black circles) and RRc (white circles) stars in the observed Fornax fields.}
\label{fig:cmd}
\end{figure}

We used a list of RRL variables from \citet{bersier02} as a reference to cross-identify RRLs in our four HAWK-I fields. Other studies of RRLs in the Fornax galaxy \citep{greco07,greco09} focus on globular clusters, and as such are unconnected to our study of field RRL stars. Although \citet{greco05astroph} examined field RRLs in Fornax, they did not provide a list of observed targets, and therefore we could not include it in our study. 
\citeauthor{bersier02} field of study overlaps ours in about 11\%, and this coverage was enough to pair 77 RRL stars (15\% of \citeauthor{bersier02} sample), among which are 66 RRab and 11 RRc stars. The position of our RRLs on the $K,J-K$ color-magnitude diagram (CMD) is shown in Fig. \ref{fig:cmd}. The scatter is caused by the single-phase nature of our measurements. The $J$- and $K$-band random-phase magnitudes of the final RRL sample together with timestamps of observations, formal photometry errors from DAOPHOT, and pulsational periods from the reference list of \citet{bersier02} are presented in Table \ref{tab:rrlyr}. In case of 25 RRL stars, which were observed more than once or found in overlapping fields, we present individual observations in separate rows. For the purpose of the analysis described in the next section, these multiple measurements were averaged, which is expected to lead to a better approximation of the mean magnitude.

\section{Distance determination}
\label{sec:distdeterm}
We used the following NIR PLZ relations to determine absolute magnitudes of RRL stars in the Fornax galaxy:

\begin{eqnarray}
M_J =& -1.773 \log P + 0.190 \log Z -0.141 \\
&\mbox{\citep{catelan04}} \nonumber
\end{eqnarray}
\begin{eqnarray}
M_K =& -2.353 \log P + 0.175 \log Z -0.1597 \\
&\mbox{\citep{catelan04}} \nonumber
\end{eqnarray}
\begin{eqnarray}
M_K =& -2.101 \log P + 0.231 \,\mathrm{[Fe/H]} -0.77 \\
&\mbox{\citep{bono03}} \nonumber
\end{eqnarray}
\begin{eqnarray}
M_K =& -2.138 \log P + 0.08 \,\mathrm{[Fe/H]} -1.07 \\
&\mbox{\citep{sollima08}} \nonumber
\end{eqnarray}

\movetabledown=1.8in
\begin{rotatetable}
\begin{deluxetable*}{ccccccCCCCCC}
\tablewidth{0pc}
\colnumbers
\tablecaption{
\label{tab:rrlyr}
Cross-identification and main characteristics of 77 RR Lyrae stars analyzed in this paper. In case of 25 RRL stars observed more than once, individual observations are placed in separate rows.}
\tablehead{
\colhead{Star ID\tablenotemark{a}} & 
\colhead{HAWK-I field} & 
\colhead{R.A.} & 
\colhead{DEC} & 
\colhead{Period\tablenotemark{a}} & 
\colhead{Type\tablenotemark{a}} & 
\colhead{$J$} &
\colhead{$\sigma_J$} & 
\colhead{$K$} &
\colhead{$\sigma_K$} &
\colhead{JD of} &
\colhead{JD of}
\\
\colhead{} & 
\colhead{} & 
\colhead{(J2000.0)} & 
\colhead{(J2000.0)} & 
\colhead{(d)} & 
\colhead{} &  
\colhead{(mag)} & 
\colhead{(mag)} & 
\colhead{(mag)} & 
\colhead{(mag)} &
\colhead{$J$ exposure} &
\colhead{$K$ exposure}
}
\startdata
FBWJ023927.1-342427 & For-III & 02:39:27.27 & -34:24:25.4 & 0.56955 & ab & 20.261 & 0.017 & 20.073 & 0.028 & 2454747.70985 & 2454747.72707 \\
FBWJ023927.2-342220 & For-III & 02:39:27.42 & -34:22:18.9 & 0.35446 & c & 20.362 & 0.016 & 20.192 & 0.032 & 2454747.70985 & 2454747.72707 \\
FBWJ023927.6-342124 & For-III & 02:39:27.88 & -34:21:23.0 & 0.60209 & ab & 20.396 & 0.019 & 20.105 & 0.040 & 2454747.70985 & 2454747.72707 \\
FBWJ023928.1-342655 & For-III & 02:39:28.28 & -34:26:53.9 & 0.36294 & c & 20.610 & 0.021 & 20.317 & 0.040 & 2454747.70985 & 2454747.72707 \\
FBWJ023939.1-342741 & For-III & 02:39:39.34 & -34:27:39.8 & 0.53957 & ab & 20.373 & 0.026 & 20.101 & 0.042 & 2454747.70985 & 2454747.72707 \\
FBWJ023939.3-342711 & For-III & 02:39:39.51 & -34:27:09.9 & 0.56782 & ab & 20.342 & 0.023 & 20.134 & 0.044 & 2454747.70985 & 2454747.72707 \\
FBWJ023942.1-342712 & For-III & 02:39:42.20 & -34:27:11.5 & 0.60813 & ab & 20.216 & 0.023 & 20.069 & 0.040 & 2454747.70985 & 2454747.72707 \\
FBWJ023943.3-342738 & For-III & 02:39:43.43 & -34:27:36.9 & 0.56517 & ab & 20.386 & 0.017 & 20.146 & 0.030 & 2454747.70985 & 2454747.72707 \\
FBWJ023944.6-342514 & For-III & 02:39:44.74 & -34:25:12.8 & 0.57117 & ab & 20.511 & 0.028 & 20.212 & 0.049 & 2454747.70985 & 2454747.72707 \\
FBWJ023945.7-342522 & For-III & 02:39:45.90 & -34:25:20.7 & 0.66012 & ab & 20.436 & 0.021 & 20.086 & 0.034 & 2454747.70985 & 2454747.72707 \\
FBWJ023947.3-342836 & For-III & 02:39:47.43 & -34:28:34.6 & 0.54003 & ab & 20.656 & 0.023 & 20.459 & 0.041 & 2454747.70985 & 2454747.72707 \\
FBWJ023949.6-342358 & For-III & 02:39:49.72 & -34:23:57.0 & 0.33563 & c & 20.446 & 0.028 & 20.278 & 0.032 & 2454747.70985 & 2454747.72707 \\
FBWJ023950.3-342320 & For-III & 02:39:50.48 & -34:23:19.1 & 0.37304 & c & 20.280 & 0.017 & 20.126 & 0.031 & 2454747.70985 & 2454747.72707 \\
FBWJ023955.0-343301 & For-II & 02:39:55.12 & -34:33:01.9 & 0.62293 & ab & 20.425 & 0.016 & 20.118 & 0.042 & 2454762.81877 & 2454762.83596 \\
FBWJ023955.4-342317 & For-III & 02:39:55.60 & -34:23:16.8 & 0.61668 & ab & 20.311 & 0.017 & 20.104 & 0.027 & 2454747.70985 & 2454747.72707 \\
\enddata
\tablenotetext{a}{Information extracted from \citet{bersier02}.}
\tablecomments{The columns contain: (1) identification of \citet{bersier02}, (2) locus of the star in our HAWK-I fields, (3) right ascension, (4) declination, (5) pulsational periods, (6) type of RRL star, (7) random-phase $J$ magnitudes with (8) their uncertainties, (9) random-phase $K$ magnitudes with (10) their  uncertainties, and the Julian Date of observations in (11) $J$ and (12) $K$ band.}
\tablecomments{Table is published in its entirety in the machine-readable format. A portion is shown here for guidance regarding its form and content.}
\end{deluxetable*}
\end{rotatetable}

\newpage
\clearpage

\begin{figure}
\includegraphics[width=\columnwidth]{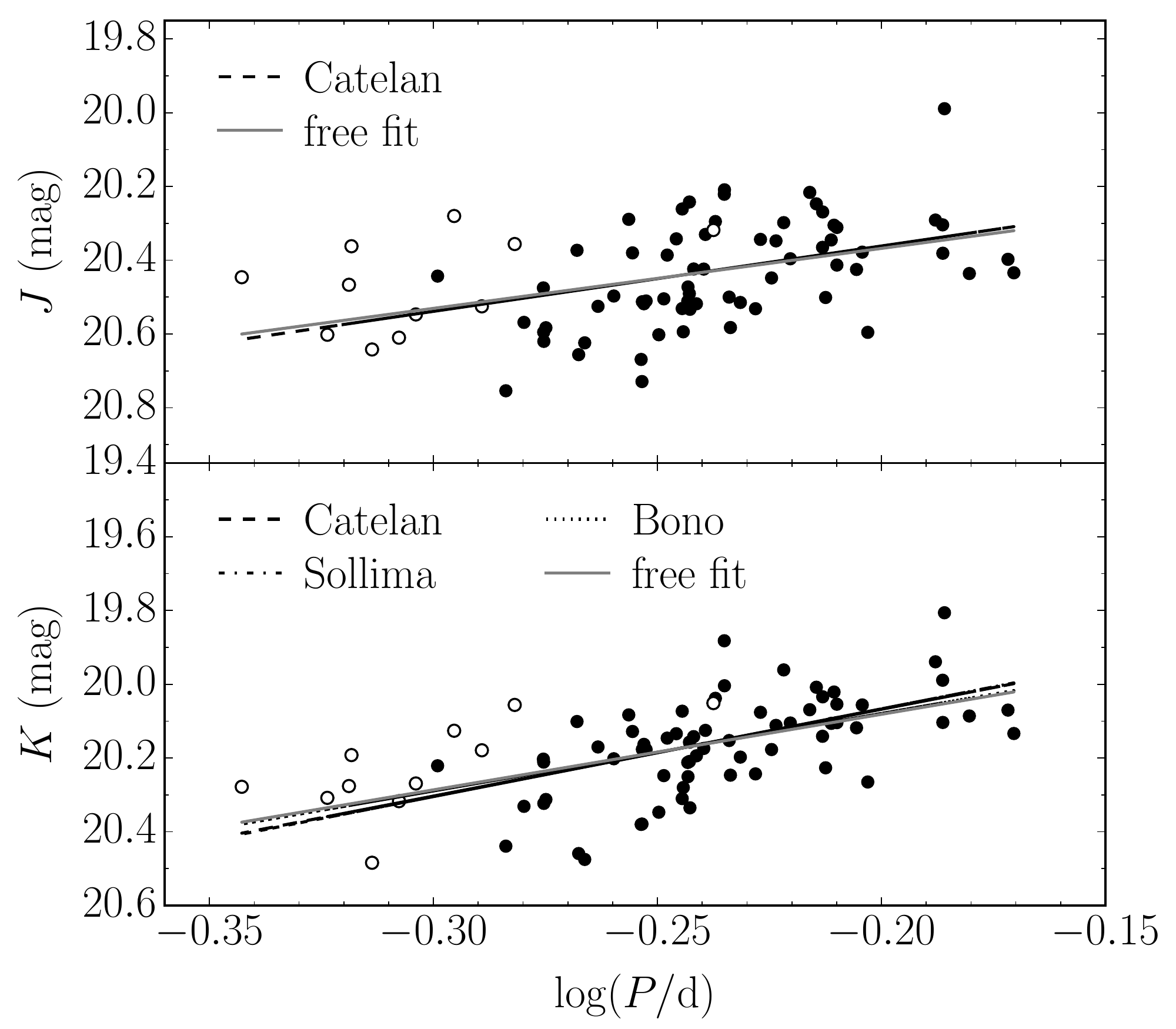}
\caption{
Period-luminosity relations for $J$- and $K$-bands defined by combined sample of 77 RRL stars (66 RRab + 11 ``fundamentalized'' RRc) observed in the Fornax galaxy, plotted along with the best fitted lines. The slopes of the fits were adopted from chosen theoretical and empirical calibrations, and the zero points were determined from our data. Filled and open circles denote RRab and RRc stars, respectively. The observed scatter is related mostly to the single-phase nature of our photometry. Magnitude error bars are within the size of circles.}
\label{fig:zp}
\end{figure}

We recall that the calibration of \citet{sollima08} was constructed for the 2MASS photometric system\footnote{Note that in the following we will use $K$ notation for 2MASS $K_{\rm s}$ passband.}, while the calibrations of \citet{bono03} and \citet{catelan04} are valid for the Bessel, Brett and Glass system (BBG).
We therefore transformed our measurements (calibrated onto UKIRT photometric system) to BBG and 2MASS systems using the transformations of \citet{carpenter01}.
The Eqs. (1)--(4) are valid for RRL fundamental periods, so in order to include first-overtone pulsators their periods were ``fundamentalized'' by adding a value of 0.127 to their logarithmic periods.
Apparent magnitudes from our dataset of 77 RRLs were individually corrected for the foreground reddening and plotted against their logarithmic periods (Fig. \ref{fig:zp}). Both periods and reddening values were extracted from \citet{bersier02}. The average reddening of our sample is $E(B-V) = 0.021$ mag, which translates to average extinction values $A_J = 0.019$ mag and $A_K = 0.008$ mag, by applying the \citet{cardelli89} reddening law $R_V = 3.1$. Following \citet{mcnamara11}, we treated internal reddening in Fornax as insignificant due to the fact that old and low-mass objects, such as RRL stars, are very unlikely to be significantly obscured by the interstellar gas from dynamic evolution or mass outflow.

The linear least-square fit performed on our sample yielded the slopes of $-1.63 \pm 0.39$ and $-2.05 \pm 0.34$ for the $J$- and $K$-band, respectively. These values agree within errors with the literature slopes from Eqs. (1)$-$(4), as can be seen in Fig. \ref{fig:zp}. In order to reduce the uncertainty of the free parameter---the zero point of the calibration---we took advantage of this consistency and performed another least-square fit, this time with the slopes intentionally fixed on their literature values. Lastly, the metallicity of Fornax [Fe/H] = $-1.6 \pm 0.2$~dex \citep{bersier02} was applied to Eqs. (3) and (4), while for Eqs. (1) and (2) the metallicity term was transformed into $\log Z$ using Eq. (9) of \citet{catelan04}.

True distance moduli calculated from Eqs. (1)$-$(4), together with associated uncertainties, are summarized in Table \ref{tab:truedist}. As the final distance modulus to the Fornax galaxy we adopt the average of these values, coupled with the largest of statistical errors. This yields a distance modulus of $20.818 \pm 0.015$ mag. Systematic errors are discussed in the following section.

\begin{deluxetable}{cccccc}
\tablewidth{0pc}
\tablecaption{
\label{tab:truedist}
True distance moduli derived from different calibrations.}
\tablehead{ \colhead{Filter} & \colhead{$K$} & \colhead{$K$} & \colhead{$K$} & \colhead{$J$} \\
\hline
Calibration &
Sollima &
Bono &
Catelan &
Catelan}
\startdata
$(m-M)_0$ & 20.787 & 20.834	& 20.820 & 20.837 \\
Statistical error	& \phn 0.013 	& \phn 0.013 	& \phn 0.013 	& \phn 0.015 \\
Systematic error 	& \phn 0.116	& \phn 0.084	& \phn 0.076	& \phn 0.083 \\
\enddata
\end{deluxetable}

\newpage
\section{Discussion}

The results in Table \ref{tab:truedist} demonstrate that the true distance moduli obtained from theoretical and empirical NIR PLZ relations for RRL stars are consistent to within their quoted errors. Among them, results from three theoretical calibrations \citep{bono03,catelan04} are virtually identical, while the calibration of \citet{sollima08} diverges by as much as 0.06 mag. The same remark was made before by e.g \citet{karczmarek15} and \citet{szewczyk09} in the context of distance determinations to the Carina galaxy and the SMC, together with the conclusion that this difference is not significant, taking into account all the uncertainties, which affect the entire process of calculating distance moduli.

However, the reappearance of this particular discrepancy between empirical and theoretical distance moduli may reveal some calibration difficulties that are still to overcome in order to achieve a full consistency of results.
One of possible issues is the underestimation of the effect of metallicity in Eq. (4). \citet{muraveva15} reported a similarly low value of metallicity coefficient based on RRL stars in the LMC, and explained that the dependence on the metallicity can be decreased if it is derived from a stellar sample characterized by narrow metallicity range. Since both metallicity and distance vary from cluster to cluster, the empirical calibration suffers from degeneracy between the metallicity coefficient and the zero point, as no constraints can be assessed for metallicity without the prior knowledge of the distance and reddening. The metallicity coefficient in Eq. (4) was derived by \citet{sollima06} based on a sample of 5 datapoints---globular clusters and RR Lyrae variable itself---with known metallicities and distances, but also large uncertainties associated with them, which could have hampered the result. The uncertainty of the zero point of Sollima's calibration was based on the prototype of this class of variable stars, RR~Lyrae itself, whose distance modulus was measured via trigonometric parallax with Hubble Space Telescope \citep{benedict02}, and equals 0.11 mag. Thanks to the first Gaia data release \citep{gaia17} this value has already been reduced to 0.09 mag and by the end of the mission is expected to reach 0.02 mag, which translates to unprecedented 0.1\% precision in distance determination \citep{gould17}. Equipped with such a precise tool, we will be able to determine distance moduli from NIR PLZ relations of RRLs at the precision level of 3\%.

More subtle effects, like helium abundances or bolometric corrections \citep{catelan04}, can also slightly affect distance calibrations by altering the absolute magnitudes of RRL stars. This was the case of \citet{sollima08}, who recalculated the zero point of their earlier PLZ relation \citep{sollima06} incorporating a new bolometric correction for the $K$-band, and resulted in 0.03 mag shift towards better coherence with theoretical calibrations.

\subsection{Reddening}
We adopted the same values of foreground reddening for each RRL star as \citet{bersier02} with an uncertainty of 0.02 mag. These values are on average 0.008 mag smaller that the mean reddening $E(B-V) = 0.03 \pm 0.02$ mag towards Fornax adopted by \citet{pietrzynski09}. The difference is however negligible, being smaller than the reddening uncertainty. If we adopted the same reddening value as \citet{pietrzynski09} for all stars in our sample, the distance moduli would be smaller by 0.004 mag and 0.001 mag in the $J$- and $K$-band, respectively. This exercise demonstrates a very limited impact of reddening in the NIR domain.

\subsection{Metallicity}
Because of its non-trivial star formation history, it is troublesome to determine the metallicity of even the oldest Fornax stellar population. With no spectroscopic measurements of Fornax RRL stars, other methods of metallicity determination have been employed: metallicity dependence on visual RRL brightness \citep{lee90,cacciari03}, metallicity dependence on RRL pulsational period \citep{sandage93}, CMD comparison \citep{saviane00}. Moreover, the results were delivered in various metallicity scales. For example, \citet{bersier02} used Sandage's period-metallicity relation to get $\mbox{[Fe/H]}= -1.6$ dex in Butler-Blanco (BB) metallicity scale\footnote{In the metallicity range of our interest BB scale is about 0.2 dex more metal-rich than the scale of \citet{zinn84} and virtually the same as the scale of \citet{carretta97}.} (used by \citealt{sandage93}). Metallicities $\mbox{[Fe/H]} = -1.78$ dex in the BB scale \citep{greco05astroph} and $\mbox{[Fe/H]} = -1.8$ dex in the \citet{zinn84} scale \citep{rizzi07} for Fornax field RRL stars are also present in the literature. 
In this study, we follow \citet{bersier02} and adopt the average metallicity $\mbox{[Fe/H]}= -1.6 \pm 0.2$ dex. The large uncertainty addresses the inconclusive metallicity of Fornax old population.
In addition to the large metallicity dispersion of Fornax RRLs, the metallicity scales in Eqs. (1)$-$(3) are not uniquely defined. In the worst case scenario, the systematic differences between scales can reach 0.2 dex \citep{layden94}. This would constitute a maximum systematic error of 0.046 mag, which we include to the final error budget.

\subsection{Alteration of the zero point by single-phase measurements}
Here we explore the possibility that at the moment of observations a significant number of RRL stars from our sample could have brightnesses above (below) their mean magnitudes. Since we approximated mean magnitudes by single-phase magnitudes, these stars would bias the zero point of the distance modulus fit towards higher (lower) values. In order to evaluate this error, we generated 77 RRL light curves, drew from each light curve a single data point which served as an approximation of the mean brightness, and calculated distance modulus in the same manner as described in Section \ref{sec:distdeterm}. Next, we repeated this procedure but this time we took the mean brightness of each star from its complete light curve. We compared the differences in calculated distance moduli for 1000 sets and concluded that the mean offset between distances derived from complete light curves and from random-phase measurements is $0.002 \pm 0.01$~mag. Therefore we adopt 0.01 mag as additional source of the systematic error. We recall that in the case of stars with a couple of observation points, we took a straight average of random-phase magnitudes, which led to a better approximation of the mean magnitudes for these stars, and to a reduction in the uncertainty of the zero point fit.

\subsection{Uncertainties}
All sources of errors which affected to any extent our results are summarized below in two groups: systematic and statistical errors. The statistical error (expressed as the standard deviation from the fit) contributes in values of 0.015~mag ($J$) and 0.013~mag ($K$) to the overall distance error, and accounts for: 
(i) the intrinsic scatter caused by single-phase measurements,
(ii) the photometric error on each observation, and
(iii) the uncertainty of atmospheric extinction and color corrections.
The systematic error is associated with: 
(i) the uncertainty of adopted metallicities,
(ii) the uncertainty of the metallicity scale,
(iii) the uncertainty of the reddening correction, 
(iv) the uncertainty of the photometric zero point, 
(v) the uncertainty of literature zero point, and
(vi) the uncertainty of the zero point fit altered by single-phase measurements.
Systematic errors significantly surpass the statistical ones, and are dominated by the uncertainties of metallicity and the literature zero point.
As in the case of the statistical error, we conservatively assumed the final systematic error to be the largest one among systematic errors presented in Table \ref{tab:truedist}. Ultimately, taking into account errors of both origins, our best distance determination to the Fornax galaxy is $20.818 \pm 0.015 \mbox{ (statistical)} \pm 0.116 \mbox{ (systematic)}$ mag.

\subsection{Comparison with previous results}
In spite of continuous effort in determining the distance to the Fornax galaxy with various standard candles, still quite large spread in the results is observed (Table \ref{tab:fordist}). For example, in the visual domain the difference between measurements of \citet{bersier00} and \citet{mcnamara11} reaches almost 0.3 mag. In the NIR domain the difference of 0.1 mag is observed between distance moduli determined by \citet{gullieuszik07} and \citet{pietrzynski03,pietrzynski09}. 
Our result is closer to Pietrzy{\'n}ski's results, although agrees also with \citet{gullieuszik07} within uncertainties.

The spread in the distances presented in Table \ref{tab:fordist} can result from a variety of reasons, among which we mention three.
Different techniques employed for distance determinations could introduce a systematic shift between the results, as seen in the case of \citet{gullieuszik07} and \citet{pietrzynski03,pietrzynski09}, both deriving Fornax distance from the red clump and the tip of the red giant branch stars.\footnote{Two techniques differed in detection method of the tip of the red giant branch stars (Maximum Likelihood Algorithm versus Sobel edge-detection filter) and in the correction on population effects. The population correction, implemented by \citet{gullieuszik07} to the red clump and the tip of the red giant branch, was argued by \citet{pietrzynski03,pietrzynski09} to be negligible in the $K$-band and therefore omitted.}
Dispersion in results among studies conducted in the visual domain can be also attributed to the reddening, which is more severe in $V$ and $I$ filters than in $J$ and $K$. Finally, in the case of RRL stars, the choice of the absolute magnitude is inextricably linked to the distance determination, as discussed by \citet{mcnamara11}.
Combination of these effects could resolve the differences in distance moduli to some extent, but might not be able to provide a full coherence of all results. A thorough comparison of methodologies and values of crucial parameters (e.g. absolute magnitude, metallicity) across all distance determinations should disclose the sources of systematic differences in distance moduli, but such a study is beyond the scope of this paper.

Our result is based on four calibrations, but more NIR PLZ relations for RRL stars exist in the literature. Some are genuinely new empirical calibrations \citep{navarrete17,marconi15,braga15}, others use the coefficients standing by the logarithmic period from previous theoretical studies, or correct for the zero points \citep{benedict11,dekany13}. Our choice of calibrations was dictated by the consistency we wanted to hold with our similar previous studies of distance determinations from RRL stars \citep{pietrzynski08,szewczyk08,szewczyk09,karczmarek15}.

\section{Conclusions and summary}
We determined the distance to the Fornax Dwarf Spheroidal Galaxy from single-phase near-infrared $J$- and $K$-band observations of 77 RR~Lyrae stars. We employed four different theoretical and empirical calibrations of the near-infrared PLZ relation for RRL stars, averaged the results, and determined a reddening-free distance modulus to the Fornax galaxy of value $20.818\pm 0.015 \mbox{ (statistical)} \pm 0.116 \mbox{ (systematic)}$ mag. The method requires very modest observational effort, because NIR single-phase measurements of RRLs can successfully approximate their mean magnitudes. Our results are consistent and agree with other distance determinations obtained from a number of independent techniques and in various wavebands, especially with the red clump stars and the tip of red giant branch, examined also in the NIR domain \citep{pietrzynski03,pietrzynski09}.

Our study adds to existing distance determination a new, competitive measurement with the precision at 5\% level. The method presented in this paper is expected to reach a precision level of 3\%, once the zero point of the PLZ calibration is refined thanks to the Gaia mission. We have consistently used NIR PLZ relations for RRL stars to measure distances to five member galaxies of the Local Group (LMC, SMC, Sculptor, Carina, and Fornax) within the Araucaria Project, and proved its quality as a tool for accurate and precise distance determination, particularly for galaxies and clusters which lack young standard candles, like Cepheids. With the use of the world largest telescopes equipped with NIR detectors this method allows to measure distances up to 300 kpc.

\acknowledgments

We gratefully acknowledge the anonymous referee, whose thorough reading and insightful comments improved the content of this paper.
The research leading to these results has received funding from the European Research Council (ERC) under the European Union’s Horizon 2020 research and innovation program (grant agreement No 695099), and from the Polish National Science Centre grant MAESTRO 2012/06/A/ST9/00269. GP and WG acknowledge financial support for this work from the BASAL Centro de Astrofisica y Tecnologias Afines (CATA) PFB-06/2007. WG and MG also acknowledge financial support from the Millenium Institute of Astrophysics (MAS) of the Iniciativa Cientifica Milenio del Ministerio de Economia, Fomento y Turismo de Chile, grant IC120009.
The data presented here were acquired under ESO programmes 082.D-0123(B) and 092.D-0295(B), granting of which is gratefully acknowledged. We thank the staff of the ESO Paranal and La Silla observatories for their help in obtaining the observations.
This research has made use of the 2MASS Database, the NASA's Astrophysics Data System Service, and the NASA/IPAC Extragalactic Database.
\facilities{VLT:Yepun (HAWK-I), NTT (SOFI)}
\software{Astropy \citep{astropy13}, Matplotlib \citep{matplotlib07}, Scipy \citep{scipy01}}

\bibliography{pkarczmarek}

\begin{thebibliography}{}
\expandafter\ifx\csname natexlab\endcsname\relax\def\natexlab#1{#1}\fi
\providecommand{\url}[1]{\href{#1}{#1}}

\bibitem[{{Astropy Collaboration} {et~al.}(2013){Astropy Collaboration},
  {Robitaille}, {Tollerud}, {Greenfield}, {Droettboom}, {Bray}, {Aldcroft},
  {Davis}, {Ginsburg}, {Price-Whelan}, {Kerzendorf}, {Conley}, {Crighton},
  {Barbary}, {Muna}, {Ferguson}, {Grollier}, {Parikh}, {Nair}, {Unther},
  {Deil}, {Woillez}, {Conseil}, {Kramer}, {Turner}, {Singer}, {Fox}, {Weaver},
  {Zabalza}, {Edwards}, {Azalee Bostroem}, {Burke}, {Casey}, {Crawford},
  {Dencheva}, {Ely}, {Jenness}, {Labrie}, {Lim}, {Pierfederici}, {Pontzen},
  {Ptak}, {Refsdal}, {Servillat}, \& {Streicher}}]{astropy13}
{Astropy Collaboration}, {Robitaille}, T.~P., {Tollerud}, E.~J., {et~al.} 2013,
  \aap, 558, A33

\bibitem[{{Battaglia} {et~al.}(2006){Battaglia}, {Tolstoy}, {Helmi}, {Irwin},
  {Letarte}, {Jablonka}, {Hill}, {Venn}, {Shetrone}, {Arimoto}, {Primas},
  {Kaufer}, {Francois}, {Szeifert}, {Abel}, \& {Sadakane}}]{battaglia06}
{Battaglia}, G., {Tolstoy}, E., {Helmi}, A., {et~al.} 2006, \aap, 459, 423

\bibitem[{{Benedict} {et~al.}(2002){Benedict}, {McArthur}, {Fredrick},
  {Harrison}, {Lee}, {Slesnick}, {Rhee}, {Patterson}, {Nelan}, {Jefferys}, {van
  Altena}, {Shelus}, {Franz}, {Wasserman}, {Hemenway}, {Duncombe}, {Story},
  {Whipple}, \& {Bradley}}]{benedict02}
{Benedict}, G.~F., {McArthur}, B.~E., {Fredrick}, L.~W., {et~al.} 2002, \aj,
  123, 473

\bibitem[{{Benedict} {et~al.}(2011){Benedict}, {McArthur}, {Feast}, {Barnes},
  {Harrison}, {Bean}, {Menzies}, {Chaboyer}, {Fossati}, {Nesvacil}, {Smith},
  {Kolenberg}, {Laney}, {Kochukhov}, {Nelan}, {Shulyak}, {Taylor}, \&
  {Freedman}}]{benedict11}
{Benedict}, G.~F., {McArthur}, B.~E., {Feast}, M.~W., {et~al.} 2011, \aj, 142,
  187

\bibitem[{{Bersier}(2000)}]{bersier00}
{Bersier}, D. 2000, \apjl, 543, L23

\bibitem[{{Bersier} \& {Wood}(2002)}]{bersier02}
{Bersier}, D., \& {Wood}, P.~R. 2002, \aj, 123, 840

\bibitem[{{Bono} {et~al.}(2001){Bono}, {Caputo}, {Castellani}, {Marconi}, \&
  {Storm}}]{bono01}
{Bono}, G., {Caputo}, F., {Castellani}, V., {Marconi}, M., \& {Storm}, J. 2001,
  \mnras, 326, 1183

\bibitem[{{Bono} {et~al.}(2003){Bono}, {Caputo}, {Castellani}, {Marconi},
  {Storm}, \& {Degl'Innocenti}}]{bono03}
{Bono}, G., {Caputo}, F., {Castellani}, V., {et~al.} 2003, \mnras, 344, 1097

\bibitem[{{Braga} {et~al.}(2015){Braga}, {Dall'Ora}, {Bono}, {Stetson},
  {Ferraro}, {Iannicola}, {Marengo}, {Neeley}, {Persson}, {Buonanno},
  {Coppola}, {Freedman}, {Madore}, {Marconi}, {Matsunaga}, {Monson}, {Rich},
  {Scowcroft}, \& {Seibert}}]{braga15}
{Braga}, V.~F., {Dall'Ora}, M., {Bono}, G., {et~al.} 2015, \apj, 799, 165

\bibitem[{{Buonanno} {et~al.}(1999){Buonanno}, {Corsi}, {Castellani},
  {Marconi}, {Fusi Pecci}, \& {Zinn}}]{buonanno99}
{Buonanno}, R., {Corsi}, C.~E., {Castellani}, M., {et~al.} 1999, \aj, 118, 1671

\bibitem[{{Cacciari} \& {Clementini}(2003)}]{cacciari03}
{Cacciari}, C., \& {Clementini}, G. 2003, in Lecture Notes in Physics, Berlin
  Springer Verlag, Vol. 635, Stellar Candles for the Extragalactic Distance
  Scale, ed. D.~{Alloin} \& W.~{Gieren}, 105--122

\bibitem[{{Cardelli} {et~al.}(1989){Cardelli}, {Clayton}, \&
  {Mathis}}]{cardelli89}
{Cardelli}, J.~A., {Clayton}, G.~C., \& {Mathis}, J.~S. 1989, \apj, 345, 245

\bibitem[{{Carpenter}(2001)}]{carpenter01}
{Carpenter}, J.~M. 2001, \aj, 121, 2851

\bibitem[{{Carretta} \& {Gratton}(1997)}]{carretta97}
{Carretta}, E., \& {Gratton}, R.~G. 1997, \aaps, 121, 95

\bibitem[{{Catelan} {et~al.}(2004){Catelan}, {Pritzl}, \& {Smith}}]{catelan04}
{Catelan}, M., {Pritzl}, B.~J., \& {Smith}, H.~A. 2004, \apjs, 154, 633

\bibitem[{{Ciechanowska} {et~al.}(2010){Ciechanowska}, {Pietrzy{\'n}ski},
  {Szewczyk}, {Gieren}, \& {Soszy{\'n}ski}}]{ciechanowska10}
{Ciechanowska}, A., {Pietrzy{\'n}ski}, G., {Szewczyk}, O., {Gieren}, W., \&
  {Soszy{\'n}ski}, I. 2010, \actaa, 60, 233

\bibitem[{{Coleman} {et~al.}(2004){Coleman}, {Da Costa}, {Bland-Hawthorn},
  {Mart{\'{\i}}nez-Delgado}, {Freeman}, \& {Malin}}]{coleman04}
{Coleman}, M., {Da Costa}, G.~S., {Bland-Hawthorn}, J., {et~al.} 2004, \aj,
  127, 832

\bibitem[{{de Boer} \& {Fraser}(2016)}]{deboer16}
{de Boer}, T.~J.~L., \& {Fraser}, M. 2016, \aap, 590, A35

\bibitem[{{de Boer} {et~al.}(2013){de Boer}, {Tolstoy}, {Saha}, \&
  {Olszewski}}]{deboer13}
{de Boer}, T.~J.~L., {Tolstoy}, E., {Saha}, A., \& {Olszewski}, E.~W. 2013,
  \aap, 551, A103

\bibitem[{{D\'{e}k\'{a}ny} {et~al.}(2013){D\'{e}k\'{a}ny}, {Minniti},
  {Catelan}, {Zoccali}, {Saito}, {Hempel}, \& {Gonzalez}}]{dekany13}
{D\'{e}k\'{a}ny}, I., {Minniti}, D., {Catelan}, M., {et~al.} 2013, \apjl, 776,
  L19

\bibitem[{{Di Criscienzo} {et~al.}(2004){Di Criscienzo}, {Marconi}, \&
  {Caputo}}]{dicriscienzo04}
{Di Criscienzo}, M., {Marconi}, M., \& {Caputo}, F. 2004, \apj, 612, 1092

\bibitem[{{Fernley} {et~al.}(1987){Fernley}, {Longmore}, \&
  {Jameson}}]{fernley87}
{Fernley}, J.~A., {Longmore}, A.~J., \& {Jameson}, R.~F. 1987, in Lecture Notes
  in Physics, Berlin Springer Verlag, Vol. 274, Stellar Pulsation, ed. A.~N.
  {Cox}, W.~M. {Sparks}, \& S.~G. {Starrfield}, 239

\bibitem[{{Gaia Collaboration} {et~al.}(2017){Gaia Collaboration},
  {Clementini}, {Eyer}, {Ripepi}, {Marconi}, {Muraveva}, {Garofalo}, {Sarro},
  {Palmer}, {Luri}, \& et~al.}]{gaia17}
{Gaia Collaboration}, {Clementini}, G., {Eyer}, L., {et~al.} 2017, ArXiv
  e-prints, arXiv:1705.00688

\bibitem[{{Gieren} {et~al.}(2005){Gieren}, {Pietrzy{\'n}ski}, {Soszy{\'n}ski},
  {Bresolin}, {Kudritzki}, {Minniti}, \& {Storm}}]{gieren05}
{Gieren}, W., {Pietrzy{\'n}ski}, G., {Soszy{\'n}ski}, I., {et~al.} 2005, \apj,
  628, 695

\bibitem[{{Gieren} {et~al.}(2013){Gieren}, {G{\'o}rski}, {Pietrzy{\'n}ski},
  {Konorski}, {Suchomska}, {Graczyk}, {Pilecki}, {Bresolin}, {Kudritzki},
  {Storm}, {Karczmarek}, {Gallenne}, {Calder{\'o}n}, \& {Geisler}}]{gieren13}
{Gieren}, W., {G{\'o}rski}, M., {Pietrzy{\'n}ski}, G., {et~al.} 2013, \apj,
  773, 69

\bibitem[{{G{\'o}rski} {et~al.}(2011){G{\'o}rski}, {Pietrzy{\'n}ski}, \&
  {Gieren}}]{gorski11}
{G{\'o}rski}, M., {Pietrzy{\'n}ski}, G., \& {Gieren}, W. 2011, \aj, 141, 194

\bibitem[{{Gould} \& {Kollmeier}(2017)}]{gould17}
{Gould}, A., \& {Kollmeier}, J.~A. 2017, Journal of Korean Astronomical
  Society, 50, 1

\bibitem[{{Graczyk} {et~al.}(2014){Graczyk}, {Pietrzy{\'n}ski}, {Thompson},
  {Gieren}, {Pilecki}, {Konorski}, {Udalski}, {Soszy{\'n}ski}, {Villanova},
  {G{\'o}rski}, {Suchomska}, {Karczmarek}, {Kudritzki}, {Bresolin}, \&
  {Gallenne}}]{graczyk14}
{Graczyk}, D., {Pietrzy{\'n}ski}, G., {Thompson}, I.~B., {et~al.} 2014, \apj,
  780, 59

\bibitem[{{Greco} {et~al.}(2005){Greco}, {Clementini}, {Held}, {Poretti},
  {Catelan}, {Dell'Arciprete}, {Gullieuszik}, {Maio}, {Rizzi}, {Smith},
  {Pritzl}, {Rest}, \& {De Lee}}]{greco05astroph}
{Greco}, C., {Clementini}, G., {Held}, E.~V., {et~al.} 2005, ArXiv Astrophysics
  e-prints, astro-ph/0507244

\bibitem[{{Greco} {et~al.}(2007){Greco}, {Clementini}, {Catelan}, {Held},
  {Poretti}, {Gullieuszik}, {Maio}, {Rest}, {De Lee}, {Smith}, \&
  {Pritzl}}]{greco07}
{Greco}, C., {Clementini}, G., {Catelan}, M., {et~al.} 2007, \apj, 670, 332

\bibitem[{{Greco} {et~al.}(2009){Greco}, {Clementini}, {Catelan}, {Held},
  {Poretti}, {Gullieuszik}, {Maio}, {Rest}, {DeLee}, {Smith}, \&
  {Pritzl}}]{greco09}
---. 2009, \apj, 701, 1323

\bibitem[{{Gullieuszik} {et~al.}(2007){Gullieuszik}, {Held}, {Rizzi},
  {Saviane}, {Momany}, \& {Ortolani}}]{gullieuszik07}
{Gullieuszik}, M., {Held}, E.~V., {Rizzi}, L., {et~al.} 2007, \aap, 467, 1025

\bibitem[{{Hawarden} {et~al.}(2001){Hawarden}, {Leggett}, {Letawsky},
  {Ballantyne}, \& {Casali}}]{hawarden01}
{Hawarden}, T.~G., {Leggett}, S.~K., {Letawsky}, M.~B., {Ballantyne}, D.~R., \&
  {Casali}, M.~M. 2001, \mnras, 325, 563

\bibitem[{{Hunter}(2007)}]{matplotlib07}
{Hunter}, J.~D. 2007, Computing in Science and Engineering, 9, 90

\bibitem[{{Huxor} \& {Grebel}(2015)}]{huxor15}
{Huxor}, A.~P., \& {Grebel}, E.~K. 2015, \mnras, 453, 2653

\bibitem[{{Irwin} \& {Hatzidimitriou}(1995)}]{irwin95}
{Irwin}, M., \& {Hatzidimitriou}, D. 1995, \mnras, 277, 1354

\bibitem[{Jones {et~al.}(2001--)Jones, Oliphant, Peterson, {et~al.}}]{scipy01}
Jones, E., Oliphant, T., Peterson, P., {et~al.} 2001--, {SciPy}: Open source
  scientific tools for {Python}, , , [Online; accessed 2017-08-04].
\newblock \url{http://www.scipy.org/}

\bibitem[{{Karczmarek} {et~al.}(2015){Karczmarek}, {Pietrzy{\'n}ski}, {Gieren},
  {Suchomska}, {Konorski}, {G{\'o}rski}, {Pilecki}, {Graczyk}, \&
  {Wielg{\'o}rski}}]{karczmarek15}
{Karczmarek}, P., {Pietrzy{\'n}ski}, G., {Gieren}, W., {et~al.} 2015, \aj, 150,
  90

\bibitem[{{Kissler-Patig} {et~al.}(2008){Kissler-Patig}, {Pirard}, {Casali},
  {Moorwood}, {Ageorges}, {Alves de Oliveira}, {Baksai}, {Bedin}, {Bendek},
  {Biereichel}, {Delabre}, {Dorn}, {Esteves}, {Finger}, {Gojak}, {Huster},
  {Jung}, {Kiekebush}, {Klein}, {Koch}, {Lizon}, {Mehrgan}, {Petr-Gotzens},
  {Pritchard}, {Selman}, \& {Stegmeier}}]{kissler08}
{Kissler-Patig}, M., {Pirard}, J.-F., {Casali}, M., {et~al.} 2008, \aap, 491,
  941

\bibitem[{{Layden}(1994)}]{layden94}
{Layden}, A.~C. 1994, \aj, 108, 1016

\bibitem[{{Lee} {et~al.}(1990){Lee}, {Demarque}, \& {Zinn}}]{lee90}
{Lee}, Y.-W., {Demarque}, P., \& {Zinn}, R. 1990, \apj, 350, 155

\bibitem[{{Longmore} {et~al.}(1986){Longmore}, {Fernley}, \&
  {Jameson}}]{longmore86}
{Longmore}, A.~J., {Fernley}, J.~A., \& {Jameson}, R.~F. 1986, \mnras, 220, 279

\bibitem[{{Marconi} {et~al.}(2003){Marconi}, {Caputo}, {Di Criscienzo}, \&
  {Castellani}}]{marconi03}
{Marconi}, M., {Caputo}, F., {Di Criscienzo}, M., \& {Castellani}, M. 2003,
  \apj, 596, 299

\bibitem[{{Marconi} {et~al.}(2015){Marconi}, {Coppola}, {Bono}, {Braga},
  {Pietrinferni}, {Buonanno}, {Castellani}, {Musella}, {Ripepi}, \&
  {Stellingwerf}}]{marconi15}
{Marconi}, M., {Coppola}, G., {Bono}, G., {et~al.} 2015, \apj, 808, 50

\bibitem[{{McGonegal} {et~al.}(1983){McGonegal}, {McAlary}, {McLaren}, \&
  {Madore}}]{mcgonegal83}
{McGonegal}, R., {McAlary}, C.~W., {McLaren}, R.~A., \& {Madore}, B.~F. 1983,
  \apj, 269, 641

\bibitem[{{McNamara}(2011)}]{mcnamara11}
{McNamara}, D.~H. 2011, \aj, 142, 110

\bibitem[{{Muraveva} {et~al.}(2015){Muraveva}, {Palmer}, {Clementini}, {Luri},
  {Cioni}, {Moretti}, {Marconi}, {Ripepi}, \& {Rubele}}]{muraveva15}
{Muraveva}, T., {Palmer}, M., {Clementini}, G., {et~al.} 2015, \apj, 807, 127

\bibitem[{{Nardetto} {et~al.}(2014){Nardetto}, {Storm}, {Gieren},
  {Pietrzy{\'n}ski}, \& {Poretti}}]{nardetto14}
{Nardetto}, N., {Storm}, J., {Gieren}, W., {Pietrzy{\'n}ski}, G., \& {Poretti},
  E. 2014, in IAU Symposium, Vol. 301, Precision Asteroseismology, ed. J.~A.
  {Guzik}, W.~J. {Chaplin}, G.~{Handler}, \& A.~{Pigulski}, 145--148

\bibitem[{{Navarrete} {et~al.}(2017){Navarrete}, {Catelan}, {Contreras Ramos},
  {Alonso-Garc{\'{\i}}a}, {Gran}, {D{\'e}k{\'a}ny}, \& {Minniti}}]{navarrete17}
{Navarrete}, C., {Catelan}, M., {Contreras Ramos}, R., {et~al.} 2017, ArXiv
  e-prints, arXiv:1704.03031

\bibitem[{{Pietrzy{\'n}ski} \& {Gieren}(2002)}]{pietrzynski02}
{Pietrzy{\'n}ski}, G., \& {Gieren}, W. 2002, \aj, 124, 2633

\bibitem[{{Pietrzy{\'n}ski} {et~al.}(2003){Pietrzy{\'n}ski}, {Gieren}, \&
  {Udalski}}]{pietrzynski03}
{Pietrzy{\'n}ski}, G., {Gieren}, W., \& {Udalski}, A. 2003, \aj, 125, 2494

\bibitem[{{Pietrzy{\'n}ski} {et~al.}(2009){Pietrzy{\'n}ski}, {G{\'o}rski},
  {Gieren}, {Ivanov}, {Bresolin}, \& {Kudritzki}}]{pietrzynski09}
{Pietrzy{\'n}ski}, G., {G{\'o}rski}, M., {Gieren}, W., {et~al.} 2009, \aj, 138,
  459

\bibitem[{{Pietrzy{\'n}ski} {et~al.}(2008){Pietrzy{\'n}ski}, {Gieren},
  {Szewczyk}, {Walker}, {Rizzi}, {Bresolin}, {Kudritzki}, {Nalewajko}, {Storm},
  {Dall'Ora}, \& {Ivanov}}]{pietrzynski08}
{Pietrzy{\'n}ski}, G., {Gieren}, W., {Szewczyk}, O., {et~al.} 2008, \aj, 135,
  1993

\bibitem[{{Pietrzy{\'n}ski} {et~al.}(2013{\natexlab{a}}){Pietrzy{\'n}ski},
  {Graczyk}, {Gieren}, {Thompson}, {Soszy{\'n}ski}, {Pilecki}, {Storm},
  {Konorski}, {Suchomska}, {Gallenne}, {Nardetto}, {Karczmarek}, \&
  {Gorski}}]{pietrzynski13eas}
{Pietrzy{\'n}ski}, G., {Graczyk}, D., {Gieren}, W., {et~al.}
  2013{\natexlab{a}}, EAS Publications Series, 64, 305

\bibitem[{{Pietrzy{\'n}ski} {et~al.}(2013{\natexlab{b}}){Pietrzy{\'n}ski},
  {Graczyk}, {Gieren}, {Thompson}, {Pilecki}, {Udalski}, {Soszy{\'n}ski},
  {Koz{\l}owski}, {Konorski}, {Suchomska}, {Bono}, {Moroni}, {Villanova},
  {Nardetto}, {Bresolin}, {Kudritzki}, {Storm}, {Gallenne}, {Smolec},
  {Minniti}, {Kubiak}, {Szyma{\'n}ski}, {Poleski}, {Wyrzykowski}, {Ulaczyk},
  {Pietrukowicz}, {G{\'o}rski}, \& {Karczmarek}}]{pietrzynski13nat}
---. 2013{\natexlab{b}}, \nat, 495, 76

\bibitem[{{Pont} {et~al.}(2004){Pont}, {Zinn}, {Gallart}, {Hardy}, \&
  {Winnick}}]{pont04}
{Pont}, F., {Zinn}, R., {Gallart}, C., {Hardy}, E., \& {Winnick}, R. 2004, \aj,
  127, 840

\bibitem[{{Poretti} {et~al.}(2008){Poretti}, {Clementini}, {Held}, {Greco},
  {Mateo}, {Dell'Arciprete}, {Rizzi}, {Gullieuszik}, \& {Maio}}]{poretti08}
{Poretti}, E., {Clementini}, G., {Held}, E.~V., {et~al.} 2008, \apj, 685, 947

\bibitem[{{Rizzi} {et~al.}(2007){Rizzi}, {Held}, {Saviane}, {Tully}, \&
  {Gullieuszik}}]{rizzi07}
{Rizzi}, L., {Held}, E.~V., {Saviane}, I., {Tully}, R.~B., \& {Gullieuszik}, M.
  2007, \mnras, 380, 1255

\bibitem[{{Sandage}(1993)}]{sandage93}
{Sandage}, A. 1993, \aj, 106, 687

\bibitem[{{Saviane} {et~al.}(2000){Saviane}, {Held}, \& {Bertelli}}]{saviane00}
{Saviane}, I., {Held}, E.~V., \& {Bertelli}, G. 2000, \aap, 355, 56

\bibitem[{{Sollima} {et~al.}(2008){Sollima}, {Cacciari}, {Arkharov},
  {Larionov}, {Gorshanov}, {Efimova}, \& {Piersimoni}}]{sollima08}
{Sollima}, A., {Cacciari}, C., {Arkharov}, A.~A.~H., {et~al.} 2008, \mnras,
  384, 1583

\bibitem[{{Sollima} {et~al.}(2006){Sollima}, {Cacciari}, \&
  {Valenti}}]{sollima06}
{Sollima}, A., {Cacciari}, C., \& {Valenti}, E. 2006, \mnras, 372, 1675

\bibitem[{{Stetson}(1987)}]{stetson87}
{Stetson}, P.~B. 1987, \pasp, 99, 191

\bibitem[{{Storm} {et~al.}(1992){Storm}, {Carney}, \& {Latham}}]{storm92}
{Storm}, J., {Carney}, B.~W., \& {Latham}, D.~W. 1992, \pasp, 104, 159

\bibitem[{{Szewczyk} {et~al.}(2009){Szewczyk}, {Pietrzy{\'n}ski}, {Gieren},
  {Ciechanowska}, {Bresolin}, \& {Kudritzki}}]{szewczyk09}
{Szewczyk}, O., {Pietrzy{\'n}ski}, G., {Gieren}, W., {et~al.} 2009, \aj, 138,
  1661

\bibitem[{{Szewczyk} {et~al.}(2008){Szewczyk}, {Pietrzy{\'n}ski}, {Gieren},
  {Storm}, {Walker}, {Rizzi}, {Kinemuchi}, {Bresolin}, {Kudritzki}, \&
  {Dall'Ora}}]{szewczyk08}
---. 2008, \aj, 136, 272

\bibitem[{{Tolstoy} {et~al.}(2001){Tolstoy}, {Irwin}, {Cole}, {Pasquini},
  {Gilmozzi}, \& {Gallagher}}]{tolstoy01}
{Tolstoy}, E., {Irwin}, M.~J., {Cole}, A.~A., {et~al.} 2001, \mnras, 327, 918

\bibitem[{{Urbaneja} {et~al.}(2008){Urbaneja}, {Kudritzki}, {Bresolin},
  {Przybilla}, {Gieren}, \& {Pietrzy{\'n}ski}}]{urbaneja08}
{Urbaneja}, M.~A., {Kudritzki}, R.-P., {Bresolin}, F., {et~al.} 2008, \apj,
  684, 118

\bibitem[{{Weisz} {et~al.}(2014){Weisz}, {Dolphin}, {Skillman}, {Holtzman},
  {Gilbert}, {Dalcanton}, \& {Williams}}]{weisz14}
{Weisz}, D.~R., {Dolphin}, A.~E., {Skillman}, E.~D., {et~al.} 2014, \apj, 789,
  147

\bibitem[{{Zinn} \& {West}(1984)}]{zinn84}
{Zinn}, R., \& {West}, M.~J. 1984, \apjs, 55, 45

\end{thebibliography}
\bibliographystyle{aasjournal}


\end{document}